# Online misinformation is linked to early COVID-19 vaccination hesitancy and refusal

Francesco Pierri[1,3]*, Brea L. Perry[2], Matthew R. DeVerna[3], Kai-Cheng Yang[3], Alessandro Flammini[3], Filippo Menczer[3] and John Bryden[3]

*Corresponding author: francesco.pierri@polimi.it

[1] Dipartimento di Elettronica, Informazione e Bioingegneria, Politecnico di Milano, Milano, Italy

[2] Department of Sociology, Indiana University, Bloomington, Indiana, USA

[3] Observatory on Social Media, Indiana University, Bloomington, Indiana, USA

## Abstract

Widespread uptake of vaccines is necessary to achieve herd immunity. However, uptake rates have varied across U.S. states during the first six months of the COVID-19 vaccination program. Misbeliefs may play an important role in vaccine hesitancy, and there is a need to understand relationships between misinformation, beliefs, behaviors, and health outcomes. Here we investigate the extent to which COVID-19 vaccination rates and vaccine hesitancy are associated with levels of online misinformation about vaccines. We also look for evidence of directionality from online misinformation to vaccine hesitancy. We find a negative relationship between misinformation and vaccination uptake rates. Online misinformation is also correlated with vaccine hesitancy rates taken from survey data. Associations between vaccine outcomes and misinformation remain significant when accounting for political as well as demographic and socioeconomic factors. While vaccine hesitancy is strongly associated with Republican vote share, we observe that the effect of online misinformation on hesitancy is strongest across Democratic rather than Republican counties. Granger causality analysis shows evidence for a directional relationship from online misinformation to vaccine hesitancy. Our results support a need for interventions that address misbeliefs, allowing individuals to make better-informed health decisions.

## Introduction

The COVID-19 pandemic has killed over 4.9 million people and infected 241 million worldwide as of October 2021 [1]. Vaccination is the lynchpin of the global strategy to fight the SARS-CoV-2 coronavirus [2], [3]. Surveys conducted during February and March 2021 found high levels of vaccine hesitancy with around 40-47% of American adults hesitant to take the COVID-19 vaccine [4], [5]. However, populations must reach a threshold vaccination rate to achieve herd immunity (i.e., 60-70%) [6]–[8]. Evidence of uneven distributions of vaccinations [9] raises the possibility of geographical clusters of non-vaccinated people [10]. In early July 2021, increased rates of the highly transmissible SARS-CoV-2 Delta variant were recorded in several poorly vaccinated U.S. states [9]. These localized outbreaks will preclude eradication of the virus and may exacerbate racial, ethnic, and socioeconomic health disparities.

Vaccine hesitancy covers a spectrum of intentions, from delaying vaccination to outright refusal to be vaccinated [11]. Some factors are linked to COVID-19 vaccine hesitancy, with rates in the U.S. highest among three groups: African Americans, women, and conservatives [12]. Other predictors, including education, employment, and income are also associated with hesitancy [13].

A number of studies discuss the spread of vaccine misinformation on social media [14] and argue that such campaigns have driven negative opinions about vaccines and even contributed to the resurgence of measles [15], [16]. In the COVID-19 pandemic scenario, widely shared misinformation includes false claims that vaccines genetically manipulate the population or contain microchips that interact with 5G networks [17], [18]. Exposure to online misinformation has been linked to increased health risks [19] and vaccine hesitancy [20]. Gaps remain in our understanding of how vaccine misinformation is linked to broad-scale patterns of COVID-19 vaccine uptake rates.

The Pfizer-BioNTec COVID-19 vaccine was the first to be given U.S. Food and Drug Administration approval on December 10th 2020 [21]. Since then, two other vaccines have been approved in the U.S. Initially, vaccines were selectively administered with nationwide priority being given to more vulnerable cohorts such as elderly members of the population. As vaccines have become available to the entire adult population more recently [22], adoption is driven by limits in demand rather than in supply. It is therefore important to study the variability in uptake across U.S. states and counties, as reflected in recent surveys [23], [24].

Here we study relationships between vaccine uptake, vaccine hesitancy, and online misinformation. Leveraging data from Twitter, Facebook, and the Centers for Disease Control and Prevention (CDC), we investigate how online misinformation is associated with vaccination rates and levels of vaccine hesitancy across the U.S. We also use Granger Causality analysis to investigate whether there is evidence for a directional association between misinformation and vaccine hesitancy.

## Methods

Our key independent variable is the mean percentage of vaccine-related misinformation shared via Twitter at the U.S. state or county level. We used 55 M tweets from the CoVaxxy dataset [17], which were collected between Jan 4th and March 25th using the Twitter filtered stream API using a comprehensive list of keywords related to vaccines (see Supplementary Information). We leveraged the Carmen library [29] to geolocate almost 1.67 M users residing in 50 U.S. states, and a subset of approximately 1.15 M users residing in over 1,300 counties. The larger set of users accounts for a total of 11 M shared tweets. Following a consolidated approach in the literature [25]–[28], we identified misinformation by considering tweets that contained links to news articles from a list of low-credibility websites compiled by a politically neutral third-party (see details in the Supplementary Information). We measured the prevalence of misinformation about vaccines in each region by (i) calculating the proportion of vaccine-related misinformation tweets shared by each geo-located account; and (ii) taking the average of this proportion across accounts within a specific region. The Twitter data collection was evaluated and deemed exempt from review by the Indiana University IRB (protocol 1102004860).

Our dependent variables include vaccination uptake rates at the state level and vaccine hesitancy at the state and county levels. Vaccination uptake is measured from the number of daily vaccinations administered in each state during the week of 19-25 March 2021, and measurements are derived from the CDC [9]. Vaccine hesitancy rates are based on Facebook Symptom Surveys provided by the Delphi Group [24] at Carnegie Mellon University. Vaccine hesitancy is likely to affect uptake rates, so we specify a longer time window to measure this variable, i.e., the period Jan 4th-March 25th 2021. We computed hesitancy by taking the complementary proportion of individuals "who either have already received a COVID vaccine or would definitely or probably choose to get vaccinated, if a vaccine were offered to them today." See Supplementary Information for further details.

There are no missing vaccine-hesitancy survey data at the state level. Observations are missing at the county level because Facebook survey data are available only when the number of respondents is at least 100. We use the same threshold on the minimum number of Twitter accounts geolocated in each county, resulting in a sample size of N = 548 counties.

Our multivariate regression models adjust for six potential confounding factors: percentage of the population below the poverty line, percentage aged 65+, percentage of residents in each racial and ethnic group (Asian, Black, Native American, and Hispanic; White non-Hispanic is omitted), rural-urban continuum code (RUCC, county level only), number of COVID-19 deaths per thousand, and percentage Republican vote (in 10 percent units). Other covariates, including religiosity, unemployment rate, and population density, were also considered (full list in Supplementary Table S9).

We also conduct a large number of sensitivity analyses, including different specifications of the misinformation variable (with a restricted set of keywords and different thresholds for the inclusion of Twitter accounts) as well as logged versions of misinformation (to correct positive skew). These results are presented in Supplementary Information (Tables S3-S8).

We conduct multiple regression models predicting vaccination rate and vaccine hesitancy. Both dependent variables are normally distributed, making weighted least squares regression the appropriate model. Data are observed (aggregated) at the state or county level rather than at the individual level. Analytic weights are applied to give more influence to observations calculated over larger samples. The weights are inversely proportional to the variance of an observation such that the variance of the j-th observation is assumed to be $\sigma^2/w_j$ where $w_j$ is the weight. The weights are set equal to the size of the sample from which the average is calculated. We estimate weighted regression with the aweights command in Stata 16. In addition, because counties are nested hierarchically within states, we use cluster robust standard errors to correct for lack of independence between county-level observations.

We investigate Granger causality between vaccine hesitancy and misinformation by comparing two auto-regressive models. The first considers daily vaccine hesitancy rates $x$ at time $t$ in geographical region $r$ (state or county):

$$x_{t,r} = \sum_i^n a_i x_{t-i,r} + \epsilon_{t,r},$$

where $n$ is the length of the time window. The second model adds daily misinformation rates per account as an exogenous variable $y$:

$$x_{t,r} = \sum_i^n (a_i x_{t-i,r} + b_i y_{t-i,r}) + \epsilon'_{t,r}.$$

The variable $y$ is said to be Granger causal [30], [31] on $x$ if, in statistically significant terms, it reduces the error term $\epsilon'_t$, i.e., if

$$E_{a,b} = \sum_{t,r} \epsilon_{t,r}^2 - \sum_{t,r} \epsilon'_{t,r}^2 > 0,$$

meaning that misinformation rates $y$ help forecast hesitancy rates $x$. We assume geographical regions to have equivalence and independence in terms of the way misinformation influences vaccine attitudes. Thus, we use the same parameters for $a_i$ and $b_i$ across all regions. We employ Ordinary Least Squares (using the Python statsmodels package version 0.11.1) linear regression to fit $a$ and $b$, standardizing the two variables and removing trends in the time series of each region. We select the value of the time window $n$ which maximizes $E_{a,b}$. For both counties and states, this was $n = 6$ days and we present results using this value. We also tested nearby values of $n \pm 2$ to confirm these provide similar results. We use data points with at least 1 tweet and at least 100 survey responses for every day in the time window for the specified region.

The traditional statistic used to assess the significance of Granger Causality is the F-statistic [30]. However, in our case, there are several reasons why this is not appropriate. First, we have missing time-windows in some of our regions. Second, our assumptions of equivalence and independence for regions may not be accurate. For these reasons, we use a bootstrap method to estimate the expected random distribution of $E_{a,b}$ with the time signal removed. To this end, we generate trial surrogates for $y$ by randomly shuffling the data points. With each random reshuffled trial, we can then use the same procedure to calculate the reduction in error, which we call $E^*_{a,b}$. The $p$-value of our Granger Causality analysis is then given by the proportion of trials ($N=10,000$) for which $E^*_{a,b} > E_{a,b}$. A potential issue with Granger Causality analysis is that it may detect an underlying trend. We tested for this by linearly detrending both time series before running the Granger analysis, finding similar results.

## Results

Looking across U.S. states, we observe a negative association between vaccination uptake rates and online misinformation (Pearson $R = -0.49$, $p < .001$). Investigating covariates known to be associated with vaccine uptake or hesitancy, we find that an increase in the mean amount of online misinformation is significantly associated with a decrease in daily vaccination rates per million ($b = -3518.00$, $p = .009$, Fig.1A, and see Methods and Supplementary Table S1). Political partisanship (a 10% increase in GOP vote) is also strongly associated with vaccination rate ($b = -640.32$, $p = .004$). These two factors alone explain nearly half the variation in state-level vaccination rates, and are themselves moderately correlated (Supplementary Fig. S1 and Table S1), consistent with prior research [32]. Remaining covariates are non-significant and/or collinear with other variables (i.e., have high variance inflation factors). and thus dropped for parsimony.

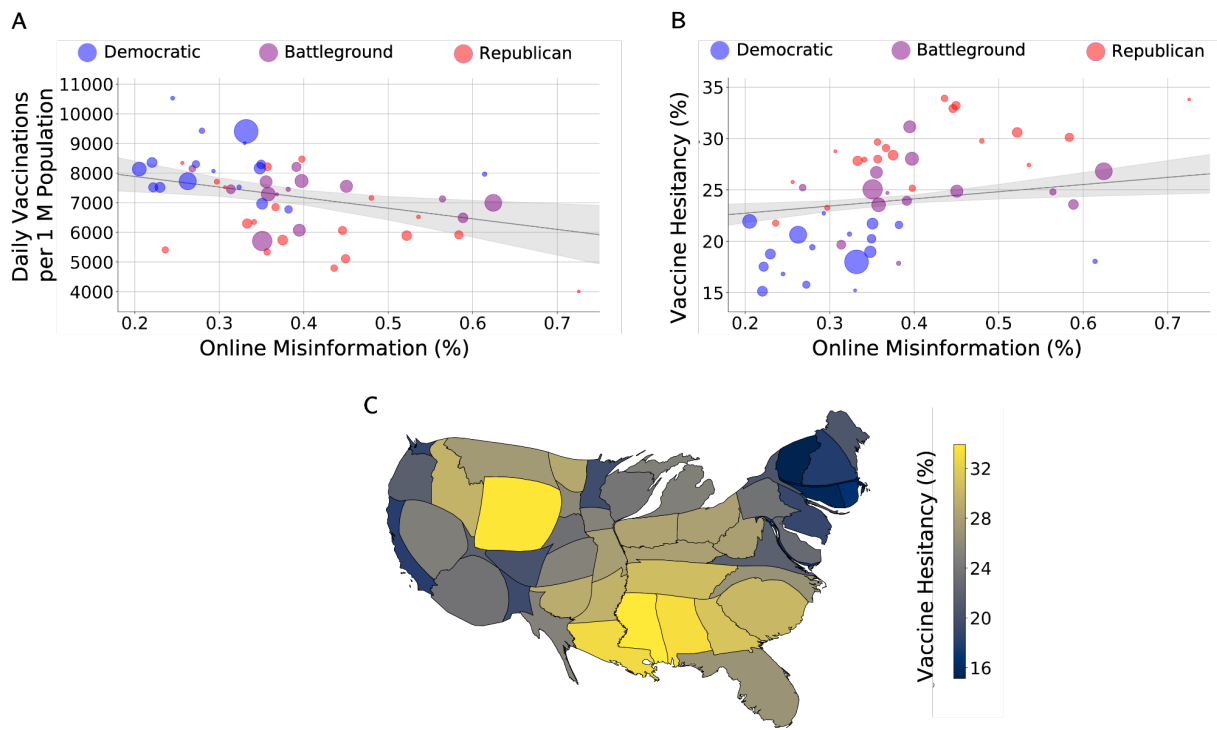

Figure 1. Online misinformation is associated with vaccination uptake and hesitancy at the state level. (A) State-level mean daily vaccinations per million population during the period from March 19 to 25, 2021, against the average proportion of vaccine misinformation tweets shared by geolocated users on Twitter during the period from Jan 4 to March 25, 2021. (B) Levels of state-wide vaccine hesitancy, computed as the fraction of individuals who would not get vaccinated according to Facebook daily surveys administered in the period from January 4 to March 25, 2021, and misinformation about vaccines shared on Twitter. Each dot represents a U.S. state and is colored according to the share of Republican voters (battleground states have a share between 45% and 55%) and sized according to population. Grey lines show the partial correlation between the two variables after adjusting for socioeconomic, demographic, and political factors in a weighted multiple linear regression model (shaded areas correspond to 95% C.I.). **(C)** Cartogram [33] (Image generated by https://go-cart.io under CC-BY license): of the U.S. in which the area of each state is proportional to the average number of misinformation links shared by geolocated users, and the color is mapped to the vaccine hesitancy rate, with lighter colors corresponding to higher hesitancy.

To investigate vaccine hesitancy, we leverage over 22 M individual responses to daily survey data provided by Facebook [24] (see Methods). Reports of vaccine hesitancy are aggregated at the state level (i.e., percent hesitant) and weighted by sample size. We find a strong negative correlation between vaccine uptake and hesitancy across U.S. states (Pearson $R = -0.71$, $p < .001$, Supplementary Fig. S1), suggesting that daily vaccination rates largely reflect demand for vaccines rather than supply. Taking into account the same set of potential confounding factors in a weighted regression model, we find a significant positive association between misinformation and state-level vaccine hesitancy ($b = 6.88$, $p = .007$), and between political partisanship and hesitancy ($b = 2.96$, $p < .001$; see Fig. 1**B** and Supplementary Fig. S1). Fig. 1**C** illustrates the state-level correlation between misinformation and hesitancy. For example, the large size and yellow color of Wyoming indicate it is the state with the highest level of misinformation and hesitancy. Among other variables, we find that the percentage of Black residents is positively related to reports of hesitancy ($b = 0.12$, $p = .001$), while the percentage of Hispanic or Latinx is negatively associated ($b = -0.07$, $p = .021$). The percentage of residents below the poverty line is also positively associated with vaccine hesitancy ($b = 0.53$, $p = .001$).

To test the robustness of these results, we also consider a more granular level of information by examining county data. Similar to previous analyses, we compute online misinformation shared by almost 1.15 M Twitter users geolocated in over 1,300 U.S. counties. We measure vaccine hesitancy rates by leveraging over 17 M daily responses to the Facebook survey for over 700 distinct counties. The total number of observations (counties) for which we are able to measure both variables is $N=548$ (see Methods). Political partisanship and misinformation are both significantly correlated with county-level vaccine hesitancy, net covariates (Supplementary Table S4 and Supplementary Fig. S2). Using a weighted multiple linear regression model, we find a significant interaction between political partisanship and misinformation. Specifically, as levels of misinformation increase, Democratic and Republican counties converge to the same level of vaccine hesitancy (Fig. 2). This may suggest the presence of a ceiling effect at around 30% of residents being vaccine hesitant (on average), with Republican counties having already reached the ceiling and thus their residents being less likely to be affected by misinformation.

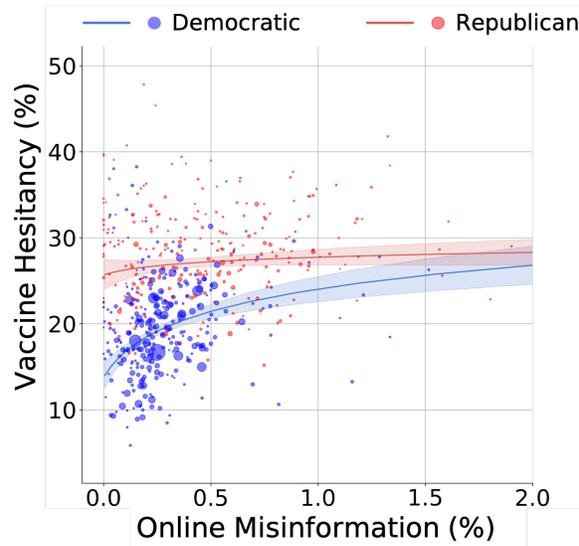

**Figure 2. Associations of online misinformation and political partisanship with vaccination hesitancy at the U.S. county level.** Each dot represents a U.S. county, with size and color indicating population size and political majority, respectively. The average proportion of misinformation shared on Twitter by geolocated users was fitted on a log scale due to non-normality (i.e., positive skew) at the county level. The two lines show predicted values of vaccine hesitancy as a function of misinformation for majority Democratic and Republican counties, adjusting for county-level confounding factors (see Methods). Shaded area corresponds to 95% C.I.

Our results so far demonstrate an association between online misinformation and vaccine hesitancy. We investigate evidence for directionality in this association by performing a Granger Causality analysis [30], [31]. We find that misinformation helps forecast vaccine hesitancy, weakly at state level ($p = .0519$) and strongly at county level ($p < .001$; see Methods and Supplementary Tables S10, S11). Analysis of the significant lagged coefficients (Supplementary Table S10) indicates that there is a lag of around 2-6 days from misinformation posted in a county to a corresponding increase in vaccine hesitancy in the same county.

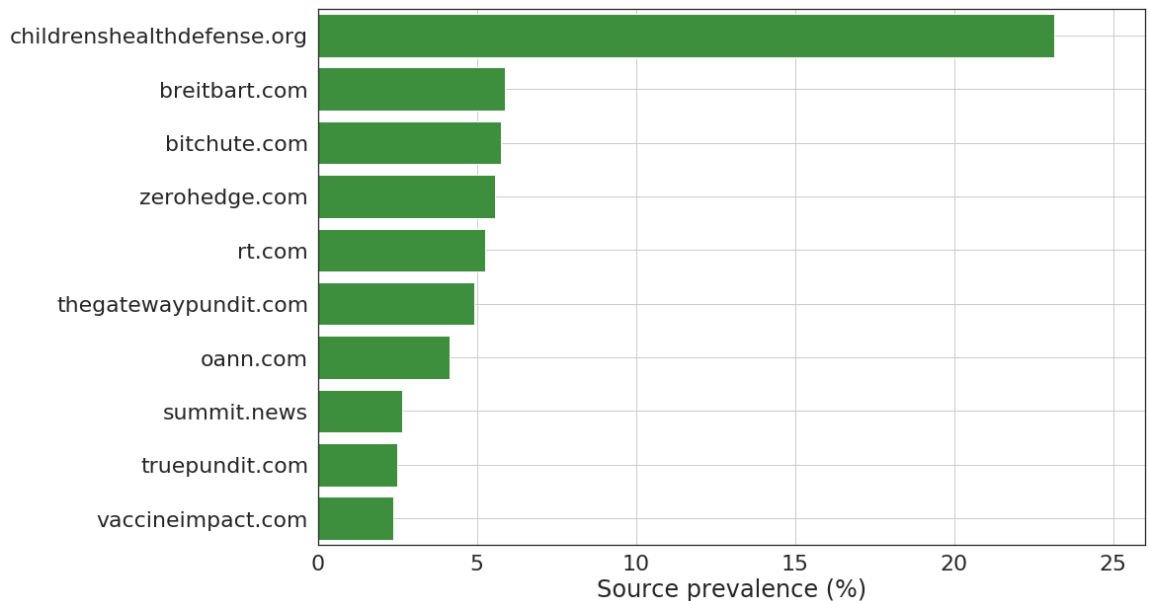

**Figure 3. Top low-credibility sources.** We considered tweets shared by users geolocated in the U.S. that link to a low-credibility source. Sources are ranked by percentage of the tweets considered.

Finally, Figure 3 shows the most shared low-credibility sources. We note the large prevalence of one particular source, Children's Health Defense, an anti-vaccination organization that has been identified as one of the main sources of misinformation on vaccines [34], [35]. We did not observe significant differences in the top sources shared in Republican vs. Democratic majority states.

# Discussion

Our results provide evidence for the problem of geographical regions with lower levels of COVID-19 vaccine uptake, which may be driven by online misinformation. Considering variability across regions with low and high levels of misinformation, the best estimates from our data predict a ~20% decrease in vaccine uptake between states, and a ~67% increase in hesitancy rates across Democratic counties, across the full range of misinformation prevalence. At these levels of vaccine uptake, the data predict SARS-CoV-2 will remain endemic in many U.S. regions. This suggests a need to counter misinformation in order to promote vaccine uptake.

An important question is whether online misinformation drives vaccine hesitancy. Our analyses alone do not demonstrate a causal relationship between misinformation and vaccine refusal. Our work is at an ecological scale and vaccine-hesitant individuals are potentially more likely to post vaccine misinformation. However, at the individual level, a recent study [20] found that exposure to online misinformation can increase vaccine hesitancy. Our work serves to provide evidence that those findings, which were obtained under controlled circumstances, scale to an ecological setting. Due to the fact that vaccine hesitancy and misinformation are socially reinforced, both ecological and individual relationships are important in demonstrating a causal link [36]. However, we are still unable to rule out confounding factors, so uncertainty remains about a causal link and further investigation is warranted.

Public opinion is very sensitive to the information ecosystem and sensational posts tend to spread widely and quickly [25]. Our results indicate that there is a geographical component to this spread, with opinions on vaccines spreading at a local scale. While social media users are not representative of the general public, existing evidence suggests that vaccine hesitancy flows across social networks [37], providing a mechanism for the lateral spread of misinformation offline among those connected directly or indirectly to misinformation spreading online. More broadly, our results provide additional insight into the effects of information diffusion on human behavior and the spread of infectious diseases [38].

A limitation of our findings is that we are not measuring the exposure, by geographical region, to misinformation on Twitter but rather the sharing activity of a subset of users. Besides, our analyses are based on data averaged over geographical regions. To account for group-level effects we present a number of sensitivity analyses, and note that our findings are consistent over two geographical scales. Our source-based approach to detect misinformation at scale might not capture the totality of misleading and harmful content related to vaccines, and many low-credibility sources publish a mixture of false and true information [39], [40]. Our results are also limited to a small period of time. Finally, other factors might also influence vaccination hesitancy levels, including accessibility to vaccines, changes in COVID-19 infection and death rates, as well as legitimate reports about vaccine safety [41].

Associations between online misinformation and detrimental offline effects, like the results presented here, call for better moderation of our information ecosystem. COVID-19 misinformation is shared overtly by known entities on major social media platforms [42]. While people have a constitutional right to free speech, it is important to maintain an environment where individuals have access to good information that benefits public health.

## Acknowledgments

We are grateful to Filipi Silva for help with the Granger Causality analysis; John Bollenbacher, Bao Tran Truong, David Axelrod, Niklas Loynes, and Christopher Torres-Lugo for their contributions to the CoVaxxy data collection and for helpful discussion; to George Turner for help with the XSEDE computer platform; and to JangDong Seo, Edwin van Leeuwen, and Francesco Scotti for insightful discussions on statistical analyses. This work is supported in part by Knight Foundation, Craig Newmark Philanthropies, the National Science Foundation (grant ACI-1548562), SSRC, and EU H2020 Periscope (grant 101016233).

## Competing interests

The authors declare no competing interests.

## Data and code availability

All measurements of vaccine uptake and vaccine hesitancy rates as well as socioeconomic, political, and demographic variables at the state and county level are publicly available in the online repository associated with this paper [43]. We also provide aggregated measurements of online misinformation shared by geolocated Twitter users. Results at the state and county level can be fully reproduced using the STATA scripts provided in the repository. Due to Twitter's terms of use and service, we can only release IDs of the tweets present in our dataset, which can be reconstructed using the Twitter API. The IDs are accessible in the public dataset associated with the CoVaxxy project [17] from the Observatory on Social Media at Indiana University.

to early COVID-19 vaccination hesitancy and refusal,'" Apr. 20, 2021. https://github.com/osome-iu/CoVaxxy-Misinfo (accessed Apr. 20, 2021).

# Supplemental Information for: Online misinformation is linked to COVID-19 vaccination hesitancy and refusal

Francesco Pierri, Brea L. Perry, Matthew R. DeVerna, Kai-Cheng Yang, Alessandro Flammini, Filippo Menczer and John Bryden

## Data collection and sources

### Twitter data

In our CoVaxxy[1] project, we collected around 55 M English-language posts about vaccines on Twitter by means of the Twitter *POST statuses/filter v1.1 API*, in the period from January 4th, 2021 to March 25th, 2021. Data collection and analysis was done using the Extreme Science and Engineering Discovery Environment (XSEDE)[2].

To define as complete a set as possible of English language keywords related to vaccines, we employed a snowball sampling methodology in December 2020[1] (see reference for full details on the data collection pipeline). The final list contains almost 80 keywords, and it is accessible in the online repository associated with the reference[3]. As a robustness test, we further perform sensitivity analyses using a restricted set of keywords ("vaccine", "vaccinate", "vaccination", "vax") which covers almost 95% of the total number of geolocated tweets. Results are equivalent to those presented in the main text and are described in the section "Sensitivity Analyses".

To match Twitter posts with US states and counties, we first identified a collection of Twitter accounts that disclosed a location in their Twitter profile. We then employed the *carmen* Python library[4] to match each location to US states and counties. We were able to match around 1.67 *M* users to 50 US states, and a subset of 1.15 *M* users to over 1,300 US counties; the larger set accounts for a total number of almost 11 *M* shared tweets.

To analyze the spread of low-credibility information, we identified all URLs shared in Twitter posts that originated from a list of low-credibility sources, following a large corpus of literature[5–9]. We employ the Iffy+ Misinfo/Disinfo list of low-credibility sources[10], which is based on information provided by the Media Bias/Fact Check website (MBFC, https://mediabiasfactcheck.com), an independent organization that reviews and rates the reliability of news sources. As defined in the related methodology, political leaning is not a factor for inclusion. The list includes sites labeled by MBFC as having a "Very Low" or "Low" factual-reporting level as well as those classified as "Questionable" or "Conspiracy-Pseudoscience". The list also includes fake-news websites flagged by BuzzFeed, FactCheck.org, PolitiFact, and Wikipedia, for a total number of 674 low-credibility sources.

Based on this list, we measure the prevalence of low-credibility information about vaccines in each region by (1) calculating the proportion of vaccine-related tweets containing URLs pointing to a low-credibility news website, for each geo-located account; and (2) taking the average of this proportion across all accounts within a specific region. We refer to this average as the state-wide (county-wide) prevalence of misinformation.

At the county level, we omit observations without vaccine hesitancy data (see next section), and we used different thresholds for the minimum number of geolocated accounts, respectively 10, 50, and 100. In the main paper, we present results when using 100 as a threshold. We provide sensitivity analyses using versions including counties with at least 10 and 50 Twitter accounts (see "Sensitivity Analyses" section). The larger threshold is likely to contain less error but also omits more counties.

## Election data

We use data provided by the MIT Election Lab to extract state-level returns for the 2020 US presidential election[11]. For counties, we use data provided by Fox News, Politico, and the New York Times. They are publicly available at https://github.com/tonmcg/US_County_Level_Election_Results_08-20.

## Vaccine hesitancy data

To compute vaccine hesitancy rates in each state (county), we leverage daily COVID-19 Symptom Surveys produced by the Delphi Group at Carnegie Mellon University[12]. These surveys are voluntarily answered by a random sample of users on Facebook (total reported sample size $N = 22{,}128{,}855$). Within the Vaccination Indicators of the survey, we extract the estimated percentage of respondents (for each state/county) "who either have already received a COVID vaccine or would definitely or probably choose to get vaccinated, if a vaccine were offered to them today." Results are available daily, for all 50 US states and for 764 US counties. We compute state-wide (county-wide) vaccine hesitancy rates by taking the proportion of negative responses in the period from January 4th to March 25th.

## Vaccine uptake data

Vaccination uptake statistics are derived from the Centers for Disease Control and Prevention (CDC) dataset (https://covid.cdc.gov/covid-data-tracker/#vaccinations). Doses monitored for each state include those administered in jurisdictional partner clinics, retail pharmacies, long-term care facilities, Federal Emergency Management Agency partner sites, Health Resources and Services Administration partner sites, and federal facilities. The data have been compiled on a daily basis by *ourworldindata.org*, and we have downloaded them for the period from January 12 to March 25, 2021. The data are available at https://github.com/owid/covid-19-data/tree/master/public/data/vaccinations.

## COVID-19 data

We extracted the number of COVID-19 cases and fatalities at the state and county level based on reports made by USAFacts (https://usafacts.org). In particular, we summed the number of daily confirmed COVID-19 cases and fatalities, referring to these as "recent", in the period from January 4 to March 25, 2021. We then computed the cumulative number of cases and fatalities on March 25th, referring to these as "total".

## Socioeconomic data

To include socioeconomic covariates in our regression model, we use data from the Atlas of Rural and Small-Town America (available at https://www.ers.usda.gov/data-products/atlas-of-rural-and-small-town-america/), which includes data at the state and county level from the American Community Survey (ACS), the Bureau of Labor Statistics, and other sources. We employ data last updated on July 2, 2020, which include county population estimates and annual unemployment/employment data for 2019. County-level measurements about religion are derived from surveys by the Association of Religion Data Archives (accessible at https://www.thearda.com/Archive/ChCounty.asp).

## Additional correlation results

Figures S1 and S2 present additional results about correlations between vaccine demand, vaccine hesitancy, political partisanship, and online misinformation at state and county levels.

# Main findings from regression analysis

Table S1 presents results from the weighted (Models 1 and 2) and ordinary (Models 3 and 4) least-squares regression of state-level vaccine hesitancy and vaccination rate, respectively, on covariates. As shown in Model 1, the misinformation variable and the percent of GOP voters explain nearly 80% of the variation in vaccine hesitancy at the state level. These predictors remain significant after the addition of multiple control variables (see Model 2). Misinformation and republican vote percentage explain nearly half of the variation in vaccination rate (see Model 3), and are also significantly associated with vaccination rate at the state level net of controls (see Model 4).

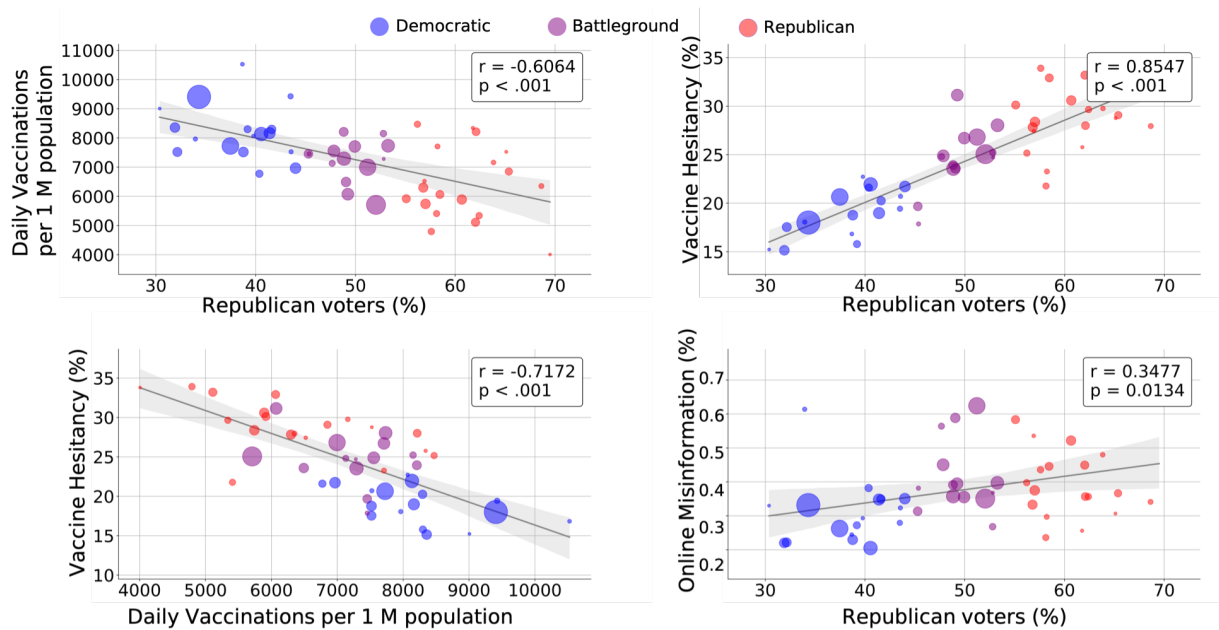

**Figure S1. Correlations between vaccine demand, vaccine hesitancy, political partisanship, and online misinformation at the state level.** Vaccine demand is computed as the mean number of daily vaccinations per million population in the period 19-25 March 2021. Vaccine hesitancy corresponds to the proportion of individuals who would not get vaccinated according to Facebook daily surveys administered in the period from January 4th to March 25th, 2021. Partisanship is measured as the percentage of Republican voters in the 2020 US Presidential elections. Online misinformation about vaccines shared on Twitter is measured during the period from Jan 4th to March 25th, 2021. Each dot represents a U.S. state, sized according to population and colored according to Republican vote share (battleground states have a share between 45% and 55%).

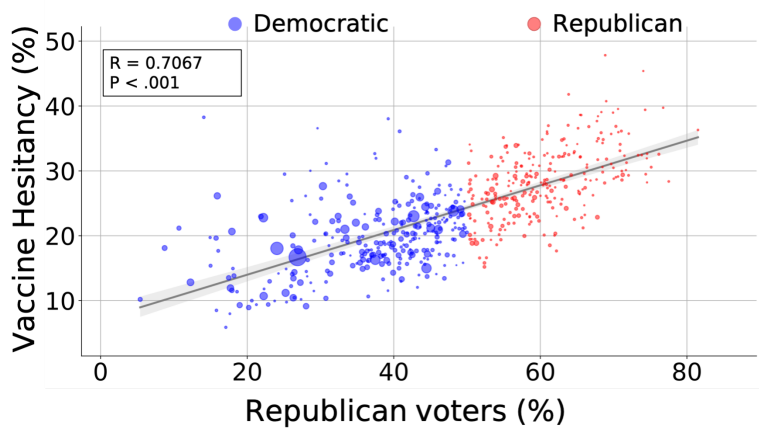

**Figure S2. Political partisanship is correlated with vaccine hesitancy at the U.S. county level.** Vaccine hesitancy corresponds to the proportion of individuals who would not get vaccinated according to Facebook daily surveys administered in the period from January 4th to March 25th, 2021. Partisanship is measured as the percentage of Republican voters in the 2020 US Presidential elections. Each dot represents a U.S. county, sized according to population and colored according to Republican vote share.

# Sensitivity analyses

We conduct a set of sensitivity analyses to ensure that our findings are robust to alternative variable and model specifications. First, we run standard diagnostics for nonlinearity, skewness, multicollinearity, and heteroskedasticity, correcting any problems we discover. Second, because the misinformation measure at the state level is slightly positively skewed, we conduct a model using a natural logarithmic transformation of mean percent misinformation. Results from these models are consistent with the main findings (Table S2). The untransformed variable has a better model fit (lower BIC). Third, because the effect of misinformation may depend on political partisanship, we test for an interaction between misinformation and the percent of GOP voters. There is no evidence of such interaction at the state level. Fourth, we rerun the above models using versions of the mean percentage of vaccine-related misinformation shared by Twitter users by considering a restricted set of keywords to gather tweets (see previous "Twitter Data" section). As shown in Table S3, findings are consistent and robust to this alternate definition of misinformation sharing.

We also conduct a similar set of sensitivity analyses at the county level. First, we test multiple versions of the misinformation variable, which is highly skewed and zero-inflated at the county level. We use the log-transformed version for the main findings due to the best model fit, but obtain significant results with the untransformed variable and very similar findings with a polynomial model that also captures the nonlinear relationship between misinformation and vaccine hesitancy. Second, we test for an interaction between misinformation and percent of GOP voters, finding that being in a majority Republican versus Democratic state moderates the association between misinformation and vaccine hesitancy (Table S4). A scatterplot of republican and democratic-leaning counties confirms the moderation finding (Fig.2 in the main manuscript). Third, we run models adding the number of tweets per county as a control variable to address variation in the volume of Twitter activity across counties. Adding this covariate did not affect results. Fourth, as at the state level, we generate versions of the vaccine misinformation variable using a restricted set of keywords. Again, these results are consistent with our main findings (Table S5). Fifth, we examine the robustness of the threshold of 100 Twitter accounts per county for inclusion in the analysis, setting thresholds of 50 and 10. These results are similar to the main findings (Tables S6 and S7), demonstrating that results are robust to different variable specifications.

To confirm the relationship between misinformation and GOP vote share, we compute a negative binomial regression model predicting mean percent information (untransformed) at the county level using percent GOP vote and a set of control variables. This multivariate analysis confirms the bivariate correlation, indicating a strong relationship between these factors net of potential confounding variables (Table S8).

Table S9 describes all the covariates considered in the regression analyses. Table S10 and S11 provide results of the OLS regression for the Granger causality analysis respectively at county and state level.

Table S1. Weighted/ordinary least squares regression of state-level percent vaccine hesitancy and daily vaccination rate per million on misinformation and covariates (N=50 states). In this and the following tables, columns correspond to different models.

|  | (1) Vaccine hesitancy b (SE) | (2) Vaccine hesitancy b (SE) | (3) Vaccination rate b (SE) | (4) Vaccination rate b (SE) |
|---|---|---|---|---|
| Mean % low credibility tweets | 8.093* (3.04) | 6.877** (2.43) | -3444.858** (1240.20) | -3518.002** (1277.08) |
| % GOP vote (10% change) | 3.996*** (0.38) | 2.960*** (0.42) | -606.567*** (140.32) | -640.319** (208.11) |
| % below poverty line |  | 0.530** (0.15) |  | 18.173 (81.84) |
| % aged 65+ |  | -0.197 (0.15) |  | 171.533 (100.14) |
| % Asian |  | 0.011 (0.07) |  | 13.213 (27.74) |
| % Black |  | 0.124** (0.04) |  | -40.491 (22.54) |
| % Hispanic |  | -0.066* (0.03) |  | 4.564 (19.71) |
| % Indigenous |  | -0.138 (0.12) |  | 71.890 (51.00) |
| COVID deaths/thousand |  | -0.221 (0.42) |  | 217.490 (262.06) |
| Constant | 1.858 (1.65) | 3.024 (2.72) | 11586.785*** (708.20) | 9126.137*** (1537.38) |
| $R^2$ | 0.797*** | 0.937*** | 0.457*** | 0.641*** |
| BIC | 225.217 | 194.454 | 836.580 | 843.252 |

Notes: Vaccine hesitancy is based on state-level means from Facebook survey data. The vaccination rate is vaccines administered per million (CDC data). For models predicting vaccine hesitancy (i.e., state means), analytic weights based on sample size are applied. Unstandardized betas and standard errors are provided. * $p < 0.05$, ** $p < 0.01$, *** $p < 0.001$

Table S2. Weighted/ordinary least squares regression of state-level percent vaccine hesitancy and daily vaccination rate per million on misinformation (logged) and covariates (N=50 states).

|  | (1) Vaccine hesitancy b (SE) | (2) Vaccine hesitancy b (SE) | (3) Vaccination rate b (SE) | (4) Vaccination rate b (SE) |
|---|---|---|---|---|
| Logged mean % low cred tweets | 4.136** (1.53) | 3.257** (1.19) | -1669.206* (636.52) | -1593.010* (660.59) |
| % GOP vote (10% change) | 3.945*** (0.38) | 2.962*** (0.42) | -601.418*** (143.03) | -676.915** (210.70) |
| % below poverty line |  | 0.515** (0.15) |  | 29.711 (83.31) |
| % aged 65+ |  | -0.158 (0.14) |  | 158.518 (101.53) |
| % Asian |  | 0.009 (0.07) |  | 8.878 (28.09) |
| % Black |  | 0.130** (0.04) |  | -42.750 (22.90) |
| % Hispanic |  | -0.062* (0.03) |  | 1.398 (19.93) |
| % Indigenous |  | -0.129 (0.12) |  | 70.503 (51.98) |
| COVID deaths/thousand |  | -0.235 (0.42) |  | 224.368 (268.26) |
| Constant | 8.318** (2.63) | 7.683 (3.90) | 8981.085*** (1015.40) | 6852.773** (2048.22) |
| $R^2$ | 0.798*** | 0.936*** | 0.448*** | 0.627*** |
| BIC | 225.049 | 194.982 | 837.352 | 845.150 |

Notes: Vaccine hesitancy is based on state-level means from Facebook survey data. The vaccination rate is actual vaccines administered per million (CDC data). For models predicting vaccine hesitancy (i.e., state means), analytic weights based on sample size are applied. Unstandardized betas and standard errors are provided. * $p < 0.05$, ** $p < 0.01$, *** $p < 0.001$

Table S3. Weighted/ordinary least squares regression of state-level percent vaccine hesitancy and daily vaccination rate per million on misinformation (restricted key words) and covariates (N=50 states).

|  | (1) Vaccine hesitancy b (SE) | (2) Vaccine hesitancy b (SE) | (3) Vaccination rate b (SE) | (4) Vaccination rate b (SE) |
|---|---|---|---|---|
| Mean % low credibility tweets | 8.320** (2.97) | 7.108** (2.37) | -3342.575** (1200.22) | -3517.510** (1236.41) |
| % GOP vote (10% change) | 3.982*** (0.37) | 2.944*** (0.41) | -611.854*** (139.58) | -648.565** (204.44) |
| % below poverty line |  | 0.517** (0.15) |  | 27.129 (81.32) |
| % aged 65+ |  | -0.206 (0.15) |  | 170.945 (99.35) |
| % Asian |  | 0.003 (0.07) |  | 16.019 (27.87) |
| % Black |  | 0.125** (0.04) |  | -42.464 (22.25) |
| % Hispanic |  | -0.065* (0.03) |  | 2.774 (19.42) |
| % Indigenous |  | -0.132 (0.12) |  | 68.678 (50.75) |
| COVID deaths/thousand |  | -0.216 (0.42) |  | 225.119 (259.70) |
| Constant | 1.841 (1.64) | 3.313 (2.71) | 11575.126*** (706.47) | 9085.430*** (1530.36) |
| $R^2$ | 0.800*** | 0.938*** | 0.457*** | 0.645*** |
| BIC | 224.530 | 193.465 | 836.543 | 842.724 |

Notes: Vaccine hesitancy is based on state-level means from Facebook survey data. The vaccination rate is actual vaccines administered per million (CDC data). For models predicting vaccine hesitancy (i.e., state means), analytic weights based on sample size are applied. Unstandardized betas and standard errors are provided. * $p < 0.05$, ** $p < 0.01$, *** $p < 0.001$

Table S4. Weighted least squares regression of county-level percent vaccine hesitancy on misinformation (logged) and covariates (N=548 counties, minimum 100 accounts/county).

|  | (1) b (SE) | (2) b (SE) | (3) b (SE) | (4) b (SE) |
|---|---|---|---|---|
| Logged mean % low credibility tweets | 1.411** (0.47) | 4.304*** (0.78) | 1.018*** (0.28) | 4.278*** (0.59) |
| % GOP vote (10% change) | 2.926*** (0.29) |  | 3.663*** (0.16) |  |
| Majority GOP state (1=GOP; 0=Dem) |  | 3.892*** (1.02) |  | 3.340*** (0.66) |
| GOP state * Logged low credibility |  | -3.585*** (0.99) |  | -3.414*** (0.76) |
| % below poverty line |  |  | 0.376*** (0.07) | 0.398*** (0.08) |
| % aged 65+ |  |  | -0.056 (0.05) | -0.091 (0.05) |
| % Asian |  |  | 0.028 (0.03) | -0.173** (0.05) |
| % Black |  |  | 0.202*** (0.02) | 0.090*** (0.03) |
| % Hispanic |  |  | 0.002 (0.02) | -0.030 (0.02) |
| % Indigenous |  |  | 0.033 (0.19) | -0.108 (0.14) |
| Rural-urban continuum code |  |  | 0.447 (0.26) | 0.617 (0.34) |
| COVID deaths/thousand |  |  | 0.547* (0.27) | 0.925** (0.29) |
| Constant | 10.227*** (1.63) | 23.668*** (1.03) | -1.535 (1.12) | 17.834*** (1.45) |
| $R^2$ | 0.500*** | 0.419*** | 0.805*** | 0.662*** |
| BIC | 3151.490 | 3240.010 | 2686.806 | 2993.820 |

Notes: Vaccine hesitancy is based on county-level means from Facebook survey data. Misinformation is measured using mean percent of low credibility tweets for counties with at least 100 Twitter accounts. Analytic weights based on Facebook survey sample size are applied, and models use cluster robust standard errors to account for counties being nested in states. Unstandardized betas and standard errors are provided. * $p < 0.05$, ** $p < 0.01$, *** $p < 0.001$

Table S5. Weighted least squares regression of county-level percent vaccine hesitancy on misinformation (logged, restricted key words) and covariates (N=548 counties, minimum 100 accounts/county).

|  | (1) b (SE) | (2) b (SE) | (3) b (SE) | (4) b (SE) |
|---|---|---|---|---|
| Logged mean % low credibility tweets | 1.510** (0.46) | 4.382*** (0.73) | 1.074*** (0.27) | 4.319*** (0.53) |
| % GOP vote (10% change) | 2.905*** (0.29) |  | 3.641*** (0.15) |  |
| Majority GOP state (1=GOP; 0=Dem) |  | 12.010*** (1.49) |  | 11.132*** (1.16) |
| GOP state * Logged low credibility |  | -3.530*** (0.94) |  | -3.392*** (0.70) |
| % below poverty line |  |  | 0.375*** (0.07) | 0.394*** (0.08) |
| % aged 65+ |  |  | -0.058 (0.05) | -0.095 (0.05) |
| % Asian |  |  | 0.028 (0.03) | -0.171** (0.05) |
| % Black |  |  | 0.202*** (0.02) | 0.091*** (0.03) |
| % Hispanic |  |  | 0.002 (0.02) | -0.030 (0.02) |
| % Indigenous |  |  | 0.038 (0.19) | -0.101 (0.13) |
| Rural-urban continuum code |  |  | 0.451 (0.26) | 0.648 (0.33) |
| COVID deaths/thousand |  |  | 0.546* (0.26) | 0.916** (0.28) |
| Constant | 6.937*** (1.14) | 13.673*** (0.95) | -3.849*** (0.93) | 7.981*** (1.29) |
| $R^2$ | 0.501*** | 0.423*** | 0.805*** | 0.665*** |
| BIC | 3136.899 | 3222.391 | 2673.021 | 2975.819 |

Notes: Vaccine hesitancy is based on county-level means from Facebook survey data. Misinformation is measured using mean percent of low credibility tweets for counties with at least 100 Twitter accounts. Analytic weights based on Facebook survey sample size are applied, and models use cluster robust standard errors to account for counties being nested in states. Unstandardized betas and standard errors are provided. * $p < 0.05$, ** $p < 0.01$, *** $p < 0.001$

Table S6. Weighted least squares regression of county-level percent vaccine hesitancy on misinformation (logged) and covariates (N=658 counties, minimum 10 accounts/county).

|  | (1) b (SE) | (2) b (SE) | (3) b (SE) | (4) b (SE) |
|---|---|---|---|---|
| Logged mean % low credibility tweets | 1.078* (0.47) | 3.252** (1.11) | 0.941*** (0.22) | 3.673*** (0.75) |
| % GOP vote (10% change) | 3.140*** (0.29) |  | 3.748*** (0.15) |  |
| Majority GOP state (1=GOP; 0=Dem) |  | 5.627*** (1.55) |  | 4.247*** (0.85) |
| GOP state * Logged low credibility |  | -2.467* (1.16) |  | -2.746** (0.84) |
| % below poverty line |  |  | 0.369*** (0.07) | 0.378*** (0.07) |
| % aged 65+ |  |  | -0.059 (0.06) | -0.114* (0.05) |
| % Asian |  |  | 0.023 (0.02) | -0.223*** (0.05) |
| % Black |  |  | 0.204*** (0.02) | 0.089*** (0.02) |
| % Hispanic |  |  | 0.002 (0.02) | -0.030 (0.02) |
| % Indigenous |  |  | -0.002 (0.12) | -0.065 (0.11) |
| Rural-urban continuum code |  |  | 0.600** (0.22) | 0.749* (0.32) |
| COVID deaths/thousand |  |  | 0.549* (0.27) | 1.054*** (0.29) |
| Constant | 9.047*** (1.65) | 22.464*** (1.58) | -2.034 (1.07) | 17.582*** (1.56) |
| $R^2$ | 0.534*** | 0.421*** | 0.812*** | 0.664*** |
| BIC | 3796.413 | 3945.657 | 3251.830 | 3639.761 |

Notes: Vaccine hesitancy is based on county-level means from Facebook survey data. Misinformation is measured using mean percent of low credibility tweets for counties with at least 10 Twitter accounts. Analytic weights based on Facebook survey sample size are applied, and models use cluster robust standard errors to account for counties being nested in states. Unstandardized betas and standard errors are provided. * $p < 0.05$, ** $p < 0.01$, *** $p < 0.001$

Table S7. Weighted least squares regression of county-level percent vaccine hesitancy on misinformation (logged) and covariates (N=628 counties, minimum 50 accounts/county).

|  | (1) b (SE) | (2) b (SE) | (3) b (SE) | (4) b (SE) |
|---|---|---|---|---|
| Logged mean % low credibility tweets | 1.347** (0.42) | 4.241*** (0.78) | 1.028*** (0.24) | 4.233*** (0.59) |
| % GOP vote (10% change) | 3.039*** (0.27) |  | 3.718*** (0.15) |  |
| Majority GOP state (1=GOP; 0=Dem) |  | 4.480*** (0.99) |  | 3.731*** (0.65) |
| GOP state * Logged low credibility |  | -3.350*** (0.90) |  | -3.236*** (0.69) |
| % below poverty line |  |  | 0.378*** (0.07) | 0.407*** (0.08) |
| % aged 65+ |  |  | -0.059 (0.06) | -0.102 (0.05) |
| % Asian |  |  | 0.030 (0.03) | -0.173** (0.05) |
| % Black |  |  | 0.202*** (0.02) | 0.087*** (0.02) |
| % Hispanic |  |  | 0.001 (0.02) | -0.034 (0.02) |
| % Indigenous |  |  | -0.008 (0.12) | -0.083 (0.10) |
| Rural-urban continuum code |  |  | 0.559* (0.23) | 0.716* (0.31) |
| COVID deaths/thousand |  |  | 0.538 (0.27) | 0.972** (0.28) |
| Constant | 9.757*** (1.48) | 23.600*** (1.03) | -1.842 (1.09) | 17.708*** (1.49) |
| $R^2$ | 0.524*** | 0.439*** | 0.809*** | 0.667*** |
| BIC | 3619.976 | 3729.469 | 3099.337 | 3453.070 |

Notes: Vaccine hesitancy is based on county-level means from Facebook survey data. Misinformation is measured using mean percent of low credibility tweets for counties with at least 50 Twitter accounts. Analytic weights based on Facebook survey sample size are applied, and models use cluster robust standard errors to account for counties being nested in states. Unstandardized betas and standard errors are provided. * $p < 0.05$, ** $p < 0.01$, *** $p < 0.001$

Table S8. Negative binomial regression of county-level misinformation on percent GOP vote and covariates (N=548 counties).

|  | b (SE) |
|---|---|
| % GOP vote (10% change) | 0.263*** |
|  | (0.04) |
| % below poverty line | -0.019* |
|  | (0.01) |
| % aged 65+ | 0.043*** |
|  | (0.01) |
| % Asian | 0.017 |
|  | (0.01) |
| % Black | 0.013*** |
|  | (0.00) |
| % Hispanic | 0.006* |
|  | (0.00) |
| % Indigenous | 0.031* |
|  | (0.02) |
| Rural-urban continuum code | -0.068 |
|  | (0.04) |
| COVID deaths/thousand | -0.098 |
|  | (0.06) |
| Constant | -2.647*** |
|  | (0.23) |
| *Wald chi-squared* | 232.330*** |
| *BIC* | 774.836 |

Notes: Misinformation is measured using mean percent of low credibility tweets for counties with at least 100 Twitter accounts. Models use cluster robust standard errors to account for counties being nested in states. Negative binomial regression is employed due to zero-inflated Poisson distribution. Unstandardized betas and standard errors are provided. * $p < 0.05$, ** $p < 0.01$, *** $p < 0.001$

Table S9. Description of covariates used during analyses.

| Stata variable | Description | Year | Source |
| --- | --- | --- | --- |
| vaxrate | Daily number of people vaccinated per million | 2021 | Centers for Disease Control and Prevention |
| lowcred | Mean percentage of low credibility shared (per user) | 2021 | Twitter API |
| loglowcred | Natural logarithm of the mean percentage of low credibility shared (per user) | 2021 | Twitter API |
| propgop | Proportion of votes for Republican candidate | 2020 | Fox News, Politico, New York Times |
| covidmortality | Total COVID 19 deaths | 2021 | Centers for Disease Control and Prevention |
| population | Census Population | 2010 | United States Census |
| vMedHHInc | Median Household Income | 2010 | United States Department of Agriculture (Atlas of Rural and Small-Town America) |
| ppoverty | Percentage of people of all ages in poverty | 2019 | United States Department of Agriculture (County-Level Datasets) |
| vPercBachelors | Percent of adults with a bachelor's degree or higher | 2015-2019 | United States Department of Agriculture (County-Level Datasets) |
| vUnemployment_rate_2019 | Unemployment rate | 2019 | United States Department of Agriculture (County-Level Datasets) |
| vTOTRATE | Rates of religious adherence per 1,000 population (200+ religions) | 2010 | Association of Religious Data Archives |
| vUnder18Pct2010 | Percentage of population age 18 years or younger | 2010 | United States Department of Agriculture (Atlas of Rural and Small-Town America) |
| vAge65AndOlderPct2010 | Percentage of population age 65 years or older | 2010 | United States Department of Agriculture (Atlas of Rural and Small-Town America) |
| vAsianNonHispPct2010 | Percentage of population Asians (Non-Hispanic) | 2010 | United States Department of Agriculture (Atlas of Rural and Small-Town America) |
| vBlackNonHispPct2010 | Percentage of population Black (Non-Hispanic) | 2010 | United States Department of Agriculture (Atlas of Rural and Small-Town America) |
| vHispanicPct2010 | Percentage of population Hispanic | 2010 | United States Department of Agriculture (Atlas of Rural and Small-Town America) |

| vNatAmNonHispPct2010 | Percentage of population Native American (Non-Hispanic) | 2010 | United States Department of Agriculture (Atlas of Rural and Small-Town America) |
|---|---|---|---|

Table S10. Ordinary Least Squares regression of lagged variates for Granger Causality analysis. (N = 610 counties).

|  | (1) | (2) | (3) | (4) | (5) | (6) |
|---|---|---|---|---|---|---|
|  | coef | std err | t | P>\|t\| | [0.025 | 0.975] |
| hesitancy t-1 | 0.8852 | 0.005 | 174.943 | 0 | 0.875 | 0.895 |
| hesitancy t-2 | 0.0039 | 0.007 | 0.571 | 0.568 | -0.009 | 0.017 |
| hesitancy t-3 | -0.0044 | 0.007 | -0.645 | 0.519 | -0.018 | 0.009 |
| hesitancy t-4 | -0.0004 | 0.007 | -0.061 | 0.951 | -0.014 | 0.013 |
| hesitancy t-5 | 0.0074 | 0.007 | 1.088 | 0.277 | -0.006 | 0.021 |
| hesitancy t-6 | -0.124 | 0.005 | -24.543 | 0 | -0.134 | -0.114 |
| misinfo t-1 | 0.006 | 0.004 | 1.362 | 0.173 | -0.003 | 0.015 |
| misinfo t-2 | 0.0087 | 0.004 | 1.972 | 0.049 | 5.36E-05 | 0.017 |
| misinfo t-3 | 0.0156 | 0.004 | 3.598 | 0 | 0.007 | 0.024 |
| misinfo t-4 | 0.0027 | 0.004 | 0.625 | 0.532 | -0.006 | 0.011 |
| misinfo t-5 | -0.0014 | 0.004 | -0.337 | 0.736 | -0.01 | 0.007 |
| misinfo t-6 | 0.0179 | 0.004 | 4.396 | 0 | 0.01 | 0.026 |
| AIC: | 56910 |  |  |  |  |  |
| R-squared (uncentered): | 0.743 |  |  |  |  |  |

Null model

|  | (1) | (2) | (3) | (4) | (5) | (6) |
|---|---|---|---|---|---|---|
|  | coef | std err | t | P>\|t\| | [0.025 | 0.975] |
| hesitancy t-1 | 0.8854 | 0.005 | 174.954 | 0 | 0.875 | 0.895 |
| hesitancy t-2 | 0.0037 | 0.007 | 0.549 | 0.583 | -0.01 | 0.017 |
| hesitancy t-3 | -0.0041 | 0.007 | -0.605 | 0.545 | -0.017 | 0.009 |
| hesitancy t-4 | -0.0005 | 0.007 | -0.079 | 0.937 | -0.014 | 0.013 |
| hesitancy t-5 | 0.0076 | 0.007 | 1.128 | 0.26 | -0.006 | 0.021 |
| hesitancy t-6 | -0.1239 | 0.005 | -24.526 | 0 | -0.134 | -0.114 |
| R-squared (uncentered): | 0.743 |  |  |  |  |  |
| AIC: | 56940 |  |  |  |  |  |

Table S11. Ordinary Least Squares regression of lagged variates for Granger Causality analysis. (N = 50 states).

|  | (1) coef | (2) std err | (3) t | (4) P>|t| | (5) [0.025 | (6) 0.975] |
|---|---|---|---|---|---|---|
| hesitancy t-1 | 0.9599 | 0.016 | 58.889 | 0 | 0.928 | 0.992 |
| hesitancy t-2 | 0.024 | 0.023 | 1.062 | 0.288 | -0.02 | 0.068 |
| hesitancy t-3 | -0.0748 | 0.022 | -3.325 | 0.001 | -0.119 | -0.031 |
| hesitancy t-4 | 0.1014 | 0.023 | 4.501 | 0 | 0.057 | 0.146 |
| hesitancy t-5 | -0.0904 | 0.023 | -3.988 | 0 | -0.135 | -0.046 |
| hesitancy t-6 | -0.0533 | 0.016 | -3.268 | 0.001 | -0.085 | -0.021 |
| misinfo t-1 | 0.0016 | 0.006 | 0.262 | 0.793 | -0.011 | 0.014 |
| misinfo t-2 | 0.021 | 0.006 | 3.351 | 0.001 | 0.009 | 0.033 |
| misinfo t-3 | 0.0018 | 0.006 | 0.295 | 0.768 | -0.01 | 0.014 |
| misinfo t-4 | -0.0161 | 0.006 | -2.603 | 0.009 | -0.028 | -0.004 |
| misinfo t-5 | 0.0133 | 0.006 | 2.153 | 0.031 | 0.001 | 0.025 |
| misinfo t-6 | 0.0003 | 0.006 | 0.044 | 0.965 | -0.012 | 0.012 |
| R-squared (uncentered): | 0.842 | | | | | |
| AIC: | 3133 | | | | | |

Null model

|  | (1) coef | (2) std err | (3) t | (4) P>|t| | (5) [0.025 | (6) 0.975] |
|---|---|---|---|---|---|---|
| hesitancy t-1 | 0.9593 | 0.016 | 58.935 | 0 | 0.927 | 0.991 |
| hesitancy t-2 | 0.0254 | 0.023 | 1.127 | 0.26 | -0.019 | 0.07 |
| hesitancy t-3 | -0.0725 | 0.023 | -3.22 | 0.001 | -0.117 | -0.028 |
| hesitancy t-4 | 0.0982 | 0.023 | 4.353 | 0 | 0.054 | 0.142 |
| hesitancy t-5 | -0.0879 | 0.023 | -3.873 | 0 | -0.132 | -0.043 |
| hesitancy t-6 | -0.0548 | 0.016 | -3.358 | 0.001 | -0.087 | -0.023 |
| R-squared (uncentered): | 0.841 | | | | | |
| AIC: | 3143 | | | | | |

# Supplementary Bibliography